# Magnetism and Nonlinear Charge Transport in $NiFe_2O_4/\gamma$-$Al_2O_3/SrTiO_3$ Heterostructure: Towards Spintronic Applications

*Amit Chanda*, Thor Hvid-Olsen, Christina Hoegfeldt, Anshu Gupta, Alessandro Palliotto, Fardin Ghaffari-Tabrizi, Maja A. Dunstan, Kasper S. Pedersen, Dae-Sung Park, Damon J. Carrad, Thomas Sand Jespersen, and Felix Trier**

Amit Chanda, Thor Hvid-Olsen, Christina Hoegfeldt, Anshu Gupta, Alessandro Palliotto, Fardin Ghaffari-Tabrizi, Dae-Sung Park, Damon J. Carrad, Thomas Sand Jespersen, and Felix Trier

Department of Energy Conversion and Storage, Technical University of Denmark

Kgs. Lyngby, 2800, Denmark

Email: amitch@dtu.dk, fetri@dtu.dk

Maja A. Dunstan, Kasper S. Pedersen

Department of Chemistry, Technical University of Denmark, Kgs. Lyngby, 2800, Denmark

**Abstract**

We present the synthesis and study of the magnetic and electronic properties of $NiFe_2O_4$/γ-$Al_2O_3$/$SrTiO_3$ heterostructure. The γ-$Al_2O_3$/$SrTiO_3$ interface hosts a high-mobility two-dimensional electron gas (2DEG) with large spin-orbit coupling, making it promising for spintronic and nano-electronic device applications if it can be coupled to a suitable source of spin currents. For efficient spin-charge conversion, a ferromagnetic insulator as a spin current source is more advantageous than a metallic/semiconducting ferromagnet. Here, we synthesize a ferrimagnetic insulating $NiFe_2O_4$(001) layer on γ-$Al_2O_3$(001)/$SrTiO_3$(001) using low-temperature reactive sputtering at 150 ℃ without compromising the mobility and charge carrier density of the 2DEG at the γ-$Al_2O_3$(001)/$SrTiO_3$(001) interface. While γ-$Al_2O_3$(8nm)/$SrTiO_3$ is nonmagnetic, $NiFe_2O_4$(52nm)/γ-$Al_2O_3$(8nm)/$SrTiO_3$ shows well-defined magnetic hysteresis loop at room temperature as well as at low temperatures. The sheet resistance of both γ-$Al_2O_3$(8nm)/$SrTiO_3$ and $NiFe_2O_4$(52nm)/γ-$Al_2O_3$(8nm)/$SrTiO_3$ exhibits metallic behavior down to cryogenic temperatures, with a low temperature upturn driven by the combined effect of Kondo-like scattering and weak antilocalization. Most importantly, $NiFe_2O_4$(52nm)/γ-$Al_2O_3$(8nm)/$SrTiO_3$ behaves as a magnetic diode at low temperatures, and its rectification performance increases significantly with increasing magnetic field strength giving rise to a robust magneto-electronic rectification effect at low temperatures, which provides a first step towards the development of nanoscale all-oxide heterostructures capable of efficient spin-charge conversion.

## 1. Introduction

A high mobility two-dimensional electron gas (2DEG) is generally formed at the interface of two insulating complex oxides,[1] *e.g.*, $SrTiO_3$ (STO) and $LaAlO_3$ and is confined within a few nanometers from the interface[2]. Such nanometric conductance confinement leads to strong interfacial Rashba spin-orbit coupling (SOC),[3] making it a very promising platform for spin-to-charge conversion via the inverse Rashba-Edelstein effect (IREE).[4–7] The efficiency of spin-charge conversion via the IREE is determined by the figure of merit, $\lambda_{IREE} = (\alpha_R \tau/\hbar)$, where $\tau = (\mu_e m_e/e)$ is the relaxation time, $\alpha_R$, $\mu_e$, $m_e$, $\hbar$, and $e$ are the Rashba coefficient, electron mobility, effective mass of electrons, reduced Planck's constant, and electronic charge, respectively. Therefore, $\lambda_{IREE}$ $(= \alpha_R \mu_e m_e/e\hbar)$ is proportional to both $\alpha_R$ and $\mu_e$ of the 2DEG.[4,6] With moderate values of $\alpha_R$ $\left(\approx 1 - 50\ \text{meVÅ}\right)$ and $\mu_e$ $(\approx 10^3\ \text{cm}^2/\text{V.s})$ at 2 K,[7,8] considerably high values of $\lambda_{\text{IEE}}$ (up to 190 nm) were obtained in STO based 2DEGs.[7–9] Notably, 2DEGs with $\mu_e \geq 10^5\ \text{cm}^2/\text{V.s}$ at 2 K have been experimentally observed at the γ-$Al_2O_3$(GAO)/STO interfaces,[2,10,11] highlighting the potential for achieving very large values of $\lambda_{IREE}$ at these interfaces, which remains to be investigated experimentally.

All past studies investigating spin-charge conversion in STO-based 2DEGs relied on conducting ferromagnets *e.g.*, $La_{0.67}Sr_{0.33}MnO_3$,[9] $Ni_{80}Fe_{20}$,[8] *etc*., as spin current sources. However, using a ferromagnetic insulator to source spin currents would be superior since conducting ferromagnets can exhibit undesirable Joule heating due to eddy currents and other parasitic effects caused by the electronic degrees of freedom, *e.g.*, anomalous Nernst effect.[12–15] However, this has not yet been experimentally realized due to the challenges in growing crystalline phases of oxide-based ferromagnetic insulators which typically requires high $O_2$ partial pressures $(\approx 10^{-2} - 10^{-1}\ \text{mbar})$ and high temperatures $(\approx 600 - 700\ ℃)$ without compromising the conducting properties of the oxygen vacancy-dominated STO-based 2DEGs.

In this context, the ferrimagnetic insulator $NiFe_2O_4$ (NFO)[16] with a high Curie temperature ($T_C \approx 850$ K)[17] could serve as a potential spin current source for efficient spin-charge conversion in the 2DEGs at the GAO/STO interfaces, as (a) both NFO and GAO possess a cubic spinel structure with ≈ 5% lattice mismatch,[2,17] (b) a crystalline phase of NFO can be grown at relatively low temperatures ($\leq 400$ °C) by reactive radio frequency (RF) sputtering with low $O_2$ partial pressure,[17–19] (c) NFO has low Gilbert damping of $\approx 10^{-4}$,[20] which facilitates long-distance magnon-propagation and hence is beneficial for efficient spin-charge conversion,[21] (d) NFO shows immense potential as efficient spin-filtering tunnel-barriers,[16,22,23] and thus, holds great promise as a highly polarized spin current source for spin-charge conversion in NFO/GAO/STO, and (e) large longitudinal spin Seebeck effect was also observed in NFO/Pt and NFO/Pd bilayers,[20,24,25] which further highlights the potential of NFO for efficient spin-charge conversion. These factors motivated us to optimize the growth and electrical properties of NFO/GAO/STO for efficient spin-charge conversion. Although determination of the spin-charge conversion efficiency of the NFO/GAO/STO heterostructure is not within the scope of this investigation, we present here a general characterization of the aforementioned heterostructure showing the successful integration of an oxide-based room temperature magnetic insulator in an oxide heterostructure hosting a moderate to high mobility 2DEG without adversely affecting its electrical properties. In this report, we demonstrate that a crystalline phase of 52 nm thick NFO can be grown by RF sputtering on GAO(8nm)/STO at a significantly lower temperature (150 °C) compared to the previous reports[17–19,26] without compromising the 2DEG's mobility and carrier density. We further show that NFO(52nm)/GAO(8nm)/STO behaves as a magnetic diode at low temperatures, and its rectification performance increases significantly with increasing magnetic field strength giving rise to a robust magneto-electronic rectification effect at low temperatures.

## 2. Results and Discussion

### 2.1. Structural and Morphological Properties

**Figure 1**(a) shows X-ray diffraction (XRD) patterns (Cu-Kα1 radiation, $\lambda$ = 1.54056 Å) for the GAO(001)/STO(001) and NFO(001)/GAO(001)/STO(001) heterostructures. Both the heterostructures exhibit a broad peak centered at $2\theta = 45.15°$ on the left side of the (002) Bragg peak of STO, which represents the (004) Bragg peak of GAO,[27,28] indicating epitaxial growth of GAO(001) on STO(001). For NFO/GAO/STO, the (004) Bragg peak of NFO(001) is visible at $2\theta = 42.11°$ (see **Fig. 1**(a)).[17,29] The estimated values of the lattice constants of GAO and NFO are $8.0259(6)$ Å and $8.576(8)$, respectively. Therefore, GAO (NFO) grows on STO (GAO) with a positive out-of-plane strain of 1.45% (2.83%), considering the $c$-lattice parameters of bulk GAO(7.911 Å) and NFO (8.334 Å).[2,17] Furthermore, a reciprocal-space-mapping performed on NFO/GAO/STO around the (002) peak of STO shows distinct peaks associated with the (004)-planes of GAO and NFO, further highlighting their single-orientation crystalline growths (see **Fig. 1**(b)). However, the X-ray rocking curves recorded on NFO/GAO/STO for the (004)-reflections of both NFO and GAO suggest poor crystallinity and significant strain-relaxation in NFO with a full-width at half-maximum (FWHM) of $(3.27 \pm 0.01)^{\circ}$ compared to GAO with an FWHM of $(0.19 \pm 0.01)^{\circ}$ (see **Fig. 1**(c)).[30,31] Evidently, in NFO/GAO/STO, Kiessig fringes in the X-ray reflectivity (XRR) spectrum associated with the thick NFO layer are superimposed on those of the thin GAO layer (see **Fig. 1**(d)). By fitting the Kiessig fringes (see **Fig. S1**), the thicknesses of GAO films in GAO/STO and NFO/GAO/STO were found to be $8.5 \pm 0.3$ nm and $8.4 \pm 0.5$ nm, respectively and that of NFO in NFO/GAO/STO was found to be $52.5 \pm 1$ nm. Moreover, while clear terraces were observed in the atomic force microscopy image of GAO(8nm)/STO (see **Fig. 1**(e)) implying crystalline and epitaxial growth of GAO, they were absent in that of NFO(52nm)/GAO(8nm)/STO, indicating poor crystallinity of NFO compared to GAO (see **Fig.**

1(f) and **Fig. S3**). Note that we have grown an additional sample of GAO/STO with thicker GAO layer using the same growth conditions as GAO(8nm)/STO the XRD pattern of which is shown in **Fig. S2**(a). As shown in **Fig. S2**(b), the thickness of this GAO film was estimated to be $52 \pm 0.2$ nm, which will serve as the control sample having roughly the same insulator thickness as the NFO(52nm)/GAO(8nm)/STO sample.

### 2.2. Magnetic Properties

The magnetic field dependent magnetization, $M(H)$ of GAO(8nm)/STO both with and without NFO was measured using a vibrating sample magnetometer associated with a Quantum Design Dynacool physical property measurement system (PPMS). While GAO(8nm)/STO shows diamagnetic behavior at 10 K (see **Fig. 2**(a)), well-defined $M(H)$ hysteresis loops were observed for NFO(52nm)/GAO(8nm)/STO at both 300 and 10 K (see **Fig. 2**(b)), demonstrating the successful synthesis of ferrimagnetic NFO. A diamagnetic linear background was subtracted from the raw $M(H)$ data of NFO(52nm)/GAO(8nm)/STO (see **Fig. S4**(a)). The values of saturation magnetization ($M_S$) and coercive field ($H_C$) for NFO(52nm)/GAO(8nm)/STO (see **Figs. S4(b)** and **(c)**) match well with previous reports of NFO grown at 400°C on $MgAl_2O_4$ substrates.[17] The temperature dependent magnetization, $M(T)$ of NFO(52nm)/GAO(8nm)/STO measured with the zero-field cooled warming (ZFCW) and field-cooled warming (FCW) protocols between 10 and 390 K in 0.5 T magnetic field is shown in **Fig. 2**(c). The bifurcation between the ZFCW and FCW $M(T)$ traces is suggestive of a glassy magnetic ground state.[32–34] In the inverse spinel structure of $NiFe_2O_4$, the octahedral sites ($O_h$) are occupied by $Ni^{2+}$ and $Fe^{3+}$ ions with a 50% occupancy and with parallel spin configuration, whereas the tetrahedral sites ($T_d$) are occupied only by the $Fe^{3+}$ ions with antiparallel spin configuration.[34–36] Oxygen deficiency in $NiFe_2O_4$ thin films destabilizes the super-exchange interaction between $Ni^{2+}$ and $Fe^{3+}$ ions and induces a canted spin configuration.[34] Such an antisymmetric exchange interaction gives rise to randomly oriented anisotropic easy axes of magnetization in exchange coupled grains and hence cluster spin-glass like magnetic ground state of $NiFe_2O_4$ thin films.[34] For deeper understanding, we measured $M(H)$ loop at 10 K for NFO(52nm)/GAO(8nm)/STO both in ZFC and FC conditions, as shown in **Fig. 2**(d). When the sample was cooled down from 390 K to 10 K in $\mu_0 H = +1$ T ($-1$ T) field, the $M(H)$ loop at 10 K shifts towards negative (positive) field directions by $\approx$ 0.275 T (0.28 T ) with respect to the ZFC $M(H)$ at 10 K, indicating the emergence of exchange-bias effect.[34] Oxygen vacancy-driven randomly oriented magnetic anisotropy axes stabilizes a cluster-spin glass state at low temperatures wherein interfacial exchange coupling between neighboring magnetic clusters gives rise to such exchange-bias effect in NFO thin films.[34] To support our magnetometry results for NFO(52nm)/GAO(8nm)/STO, we have performed X-ray photoelectron spectroscopy (XPS) measurements on this sample. From the XPS spectra of Ni $2p$ and Fe $2p$ (see **Figs. S5**(a) and (b), respectively), the oxidation states of Ni and Fe were determined as $Ni^{2+}$ and $Fe^{3+}$, respectively. The Ni: Fe ratio determined from the $3p$ regions and the Fe: O ratio from the Fe $2p$ and O $1s$ regions are both close to the nominal 1:2 ratio (see **Table S1**). Our results indicate that both $Fe^{3+}$ and $Ni^{2+}$ ions are present at the $T_d$ and $O_h$ sites, with 72% of Fe and 61% of Ni occupy the $O_h$ sites whereas the remaining 28% of Fe and 39% of Ni occupy the $T_d$ sites (see supporting information for details). Similar cationic distribution with $Ni^{2+}$ ions occupying non-equilibrium $T_d$ sites has previously been observed in oxygen-deficient NFO thin films,[34][37] which was believed to be caused by the bulk oxygen vacancies that promote migration of cations from their bulk equilibrium sites[34]. Therefore, the exchange-bias effect observed in our NFO(52nm)/GAO(8nm)/STO heterostructure can be associated with the presence of oxygen vacancies in the NFO layer. However, it is important to note that the bulk oxygen vacancy level in the 52nm NFO layer in the NFO(52nm)/GAO(8nm)/STO heterostructure is hard to quantify directly from the O 1$s$ spectra as XPS is a surface sensitive measurement technique.

## 2.3. Four-Terminal Electronic Transport

### *2.3.1. Temperature and Magnetic Field Dependent Sheet Resistance*

Temperature dependent sheet resistance, $R_s(T)$ for GAO(8nm)/STO and NFO(52nm)/GAO(8nm)/STO measured in the range: 2 K $\leq T \leq$ 300 K is shown in **Figs. 2**(e) and (f), respectively. The value of $R_s$ remains nearly unaffected by the growth of the NFO layer; it decreases from ≈ 46.6 kOhm/sq (48.06 kOhm/sq) at 300 K to ≈ 1.15 kOhm/sq (1.73 kOhm/sq) at 2 K for GAO(8nm)/STO (NFO(52nm)/GAO(8nm)/STO). For 30 K $\leq T \leq$ 300 K, $R_s$ decreases upon cooling for both GAO(8nm)/STO and NFO(52nm)/GAO(8nm)/STO, as expected for the presence of a 2DEG at the γ-$Al_2O_3$/$SrTiO_3$ interface.[38] However, $R_s$ shows an upturn below $T_K^*$ = 18 K and 27 K for GAO(8nm)/STO and NFO(52nm)/GAO(8nm)/STO, respectively which is followed by an approach to saturation in the temperature range: 2 K $\leq$ $T \leq 5$ K. Note that $R_s(T)$ for the control sample GAO(52nm)/STO also showed similar behavior: it decreases from ≈ 26.5 kOhm/sq at 300 K to ≈ 0.9 kOhm/sq at 2 K along with an upturn below $T_K^*$ = 18 K followed by an approach to saturation below 5 K (see **Fig. S11**(a)). Furthermore, a clear comparison of the upturns in $R_s(T)$ for all three samples: GAO(8nm)/STO, NFO(52nm)/GAO(8nm)/STO and GAO(52nm)/STO are presented in **Fig. S13.** Such low temperature resistive behavior of STO-based 2DEGs is generally attributed to the Kondo-like screening,[39] resulting from exchange interactions between itinerant conduction electrons of the 2DEG and localized magnetic scattering centers, *e.g.*, the localized spin-polarized $Ti^{3+}$ ions.[38,40–44] The Kondo-like scattering is experimentally manifested as a minimum in $R_s(T)$ with a $-lnT$ dependence at temperatures below the minimum[45]. We fitted the $R_s(T)$ data for both the heterostructures below 60 K using the general expression,[44,46–48]

$$R_s(T) = R_0 + R_{inel}T^n + R_{Kondo}\left(1 - \left[\frac{\ln(T/T_K)}{\sqrt{\{\ln(T/T_K)\}^2+\pi^2 S(S+1)}}\right]\right) \quad (1)$$

Here, $R_0$ represents the temperature independent residual resistance, the second term represents the inelastic scattering contribution which accounts for the combined effect of electron-electron inelastic scattering and electron-phonon scattering.[49–51] The fourth term represents the Kondo-like scattering contribution,[48] where $T_K$ is the effective Kondo temperature and $S$ is the effective spin of the magnetic scattering centers, which was fixed at $S = 0.225$ [47]. The fitting parameters for both GAO(8nm)/STO and NFO(52nm)/GAO(8nm)/STO are listed in **Table S2**, and those for GAO(52nm)/STO are listed in **Table S5**, which indicates that the Kondo-like scattering contribution is higher in NFO(52nm)/GAO(8nm)/STO compared to both GAO(8nm)/STO and GAO(52nm)/STO. Furthermore, $T_K$ is also higher in NFO(52nm)/GAO(8nm)/STO than in both GAO(8nm)/STO and GAO(52nm)/STO, which agrees with our experimental observation of the appearance of the upturn in $R_s(T)$ at a higher temperature in NFO(52nm)/GAO(8nm)/STO compared to both GAO(8nm)/STO and GAO(52nm)/STO. Such enhanced Kondo-like scattering effect in NFO(52nm)/GAO(8nm)/STO compared to both GAO(8nm)/STO and GAO(52nm)/STO is likely due to the interplay between magnetic proximity effect (MPE) and Rashba spin-orbit coupling (SOC) of the 2DEG.[52] The $R_s$ of the 2DEG and hence the associated electron scattering mechanisms depend on the Rashba SOC.[53] The MPE induced by the NFO layer influences the Rashba SOC of the 2DEG at the GAO(8nm)/STO interface,[54,55] which in turn enhances the Kondo screening effect[56]. Notably, MPE-driven enhanced Kondo effect was also observed in $EuTiO_3/KTaO_3$ based 2DEGs.[57]

In order to confirm the origin of the upturns in $R_s(T)$ for both the heterostructures, we have performed magnetic field dependent sheet resistance measurements on the NFO(52nm)/GAO(8nm)/STO heterostructure while sweeping an in-plane magnetic field between ±16 T at 6 K (see **Fig. S6**). A negative magnetoresistance was observed, which further supports the Kondo-like scattering as a possible origin of the observed resistance upturn for $T \leq 27$ K in this heterostructure.[58,59] Here, the magnetoresistance (MR) is defined as, $MR = [\{R_s(\mu_0 H) - R_s(\mu_0 H = 0)\}/R_s(\mu_0 H = 0)]$. However, in case of STO-based 2DEGs, weak antilocalization (WAL) also contributes towards charge transport at low temperatures, which can be experimentally manifested by a low-field cusp-like feature in the out-of-plane magnetoresistance.[5,38,59] At low temperatures (usually below 5 K), the conductivity of the 2DEG is enhanced due to a destructive interference effect of the coherently back-scattered electrons in the absence of external magnetic field. However, an application of external out-of-plane magnetic field rapidly suppresses this effect and the sheet resistance of the 2DEG increases sharply with increasing magnetic field in the low field regime, giving rise to a cusp-like feature near the zero field.[5,38,59] In order to clarify the contribution of weak antilocalization (WAL) effect, we have performed magnetic field dependent sheet resistance measurements on both GAO(8nm)/STO and NFO(52nm)/GAO(8nm)/STO heterostructures while sweeping an out-of-plane magnetic field between $\mu_0 H = \pm 5\ T$ at 2 K, 5 K and 10 K (see **Figs. 3**(a) and (b), respectively) using the DC resistivity option of the PPMS. Herein, a positive MR was observed. Notably, the MR at 2 K deviates from the quadratic field dependence at lower magnetic fields below $\mu_0 H = 3\ T$ and $\mu_0 H = 1\ T$ for GAO(8nm)/STO and NFO(52nm)/GAO(8nm)/STO, respectively, and shows a sharp cusp-like feature close to the zero field, indicating the presence of WAL effect.[58,59] Moreover, we observed similar cusp-like feature (weaker) out-of-plane MR for GAO(52nm)/STO (see **Fig. S12**) at 2 K, indicating the presence of WAL in this sample as well. However, this cusp-like feature becomes weaker with increasing temperature to 5 K and disappears at 10 K for both GAO(8nm)/STO and NFO(52nm)/GAO(8nm)/STO (see **Figs. 3**(a) and (b), respectively). Furthermore, we have systematically performed magnetic field dependent sheet resistance measurements on both GAO(8nm)/STO and NFO(52nm)/GAO(8nm)/STO while sweeping an out-of-plane magnetic field between $\mu_0 H = \pm 16\, T$ in the temperature range $6\, K \leq T \leq 100\, K$ in our cryostat and observed quadratic field dependence at all temperatures in this range without any low-field cusp-like feature (see **Figs. 3**(c) and (d), respectively), which clearly indicates that the contribution of WAL is significant only at and below 5 K but completely disappears for $T \geq 10\, K$. Therefore, based on these observations, we conclude that a combination of both Kondo-like scattering and WAL contributes towards the low temperature charge transport properties of the GAO(8nm)/STO and NFO(52nm)/GAO(8nm)/STO heterostructures. Notably, the absolute value of the out-of-plane MR at $\mu_0 H = 16\, T$ for GAO(8nm)/STO (NFO(52nm)/GAO(8nm)/STO) increases with temperature from ≈ 18.5 % (21 %) at $T$ = 6 K and shows a maximum value of ≈ 21 % (25 %) at $T$ = 15 K ($T$ = 25 K) and then decreases with further increasing the temperature. This is also evident from **Figs. 4**(a) and (b), wherein temperature evolution of the out-of-plane MR at $\mu_0 H = 16\, T$ are shown for GAO(8nm)/STO and NFO(52nm)/GAO(8nm)/STO, respectively. Therefore, it is apparent that the out-of-plane MR shows a maximum at $T_K^*$ for both the heterostructures.

*2.3.2. Nonlinear Hall Effect*

To ensure that introducing the NFO layer does not adversely affect the charge carrier density ($n_e$) and mobility ($\mu_e$) of the 2DEG at the GAO(8nm)/STO interface, we performed Hall measurements on both the heterostructures. **Figures 3**(e) and (f) demonstrate magnetic field dependence of Hall resistance, $R_{xy}(H)$ at different temperatures between 10 and 300 K with applied magnetic fields up to $\mu_0 H = \pm 15$ T for GAO(8nm)/STO and NFO(52nm)/GAO(8nm)/STO, respectively. As shown in **Fig. 4**(c) and **Fig. S7**, the temperature dependence of $R_{xy}$ at $\mu_0 H = +15$ T , for both NFO(52nm)/GAO(8nm)/STO and GAO(8nm)/STO, respectively shows a maximum around their respective $T_K^*$ . Most importantly, $R_{xy}(H)$ curves for both the heterostructures exhibit nearly linear field dependence above 100 K but, nonlinear field dependence appears below 100 K. Additionally, the degree of nonlinearity becomes stronger with decreasing temperature. The same behavior of the $R_{xy}(H)$ curves was observed for GAO(52nm)/STO (see **Fig. S11**(b)). Nonlinear $R_{xy}(H)$ can have the following origins: (1) ferromagnetism induced anomalous Hall effect (AHE),[60,61] (2) quantum effects (*e.g*., quantum oscillations) in presence of strong magnetic fields,[62] and (3) spatially separated multi-band conduction of the same type of carriers with different densities and mobilities.[63] We can rule out ferromagnetism-induced AHE in the case of GAO(8nm)/STO, since no signatures were detected by VSM measurements down to $T = 10$ K (see **Fig. 2**(a)). However, the degree of nonlinearity is higher in NFO(52nm)/GAO(8nm)/STO than in GAO(8nm)/STO (see **Fig. S8**), possibly due to MPE induced non-negligible AHE in NFO/GAO/STO.[60,61] Now, the quantum effects will play a significant role in $R_{xy}(H)$ when $\omega_C\tau_C > 1$; where $\omega_C = (e.\mu_0 H/m_e)$ is the cyclotron frequency and $\tau_C = (\mu_e.m_e/e)$ is the relaxation time for the scattering process, $\mu_e$ is the electron mobility, $m_e$ is the effective mass of the electron, and $e$ is the electronic charge. Therefore, the aforementioned condition can be written as, $\omega_C\tau_C = \mu_e\mu_0 H > 1$, *i.e*., $\mu_0 H > (1/\mu_e)$. In other words, the contribution of quantum effects towards nonlinear $R_{xy}(H)$ will be significant if the external magnetic field exceeds $(1/\mu_e)$. By fitting the linear low field ($-5\text{T} \leq \mu_0 H \leq +5$ T) region of the $R_{xy}(H)$ curve at 10 K with a one-band model, we obtained $\mu_e = 767$ cm$^2$V$^{-1}$s$^{-1}$ for NFO(52nm)/GAO(8nm)/STO which leads to the weak field limit of $(1/\mu_e) \approx 12.25$ T. However, the slope changes in $R_{xy}(H)$ at low temperatures for both GAO(8nm)/STO and NFO(52nm)/GAO(8nm)/STO occur around $\mu_0 H = \pm 5$ T. Therefore, the contribution of quantum effects towards the nonlinearity in $R_{xy}(H)$ curves can be ruled out for both the heterostructures. Hence, the nonlinear Hall effect observed in these heterostructures at low temperatures is primarily governed by spatially separated multi-band conduction of charge carriers with different densities and mobilities.[10]

Considering only the contribution of two-band conduction of electrons ($R_{xy}^{2e}(\mu_0 H)$), we can express $R_{xy}(\mu_0 H)$ as,[58,59,63–65]

$$R_{xy}(\mu_0 H) = R_{xy}^{2e}(\mu_0 H) = -\left(\frac{1}{e}\right)\left[\frac{\left\{\frac{n_1\mu_1^2}{1+\mu_1^2(\mu_0 H)^2}+\frac{n_2\mu_2^2}{1+\mu_2^2(\mu_0 H)^2}\right\}.(\mu_0 H)}{\left\{\frac{n_1\mu_1}{1+\mu_1^2(\mu_0 H)^2}+\frac{n_2\mu_2}{1+\mu_2^2(\mu_0 H)^2}\right\}^2+\left\{\frac{n_1\mu_1^2}{1+\mu_1^2(\mu_0 H)^2}+\frac{n_2\mu_2^2}{1+\mu_2^2(\mu_0 H)^2}\right\}^2.(\mu_0 H)^2}\right] \quad (2)$$

Here, Eqn. (2) is constrained by the expression:

$$R_s^{-1}(\mu_0 H = 0) = e(n_1\mu_1 + n_2\mu_2) \quad (3)$$

where, $n_i(i = 1, 2)$ is the carrier concentration, $\mu_i(i = 1, 2)$ is the carrier mobility of the $i^{th}$ electronic conduction band, and $R_s(\mu_0 H = 0)$ is the zero field sheet resistance. We fitted the $R_{xy}(H)$ curves for both the heterostructures using Eqns. (2) and (3) as shown in **Fig. 4**(d). The temperature evolutions of $n_i$'s and $\mu_i$'s for both the heterostructures are summarized in **Figs. 4**(e) and (f), respectively. For both the heterostructures, $n_1$ is nearly an order of magnitude higher than $n_2$ whereas $\mu_1$ is significantly lower than $\mu_2$ at all temperatures. Furthermore, for both GAO(8nm)/STO and NFO(52nm)/GAO(8nm)/STO, $n_2$ increases by nearly an order of magnitude ($\approx 10^{11}$ to $10^{12}$ cm$^{-2}$), $\mu_2$ increases relatively slowly ($\approx 400$ to $1150$ cm$^2$/(V.s)) while decreasing the temperature from 100 to 10 K, similar to the observations from $La_{0.5}Sr_{0.5}TiO_3/SrTiO_3$ heterostructures.[63] Most importantly, both $n_i$'s and $\mu_i$'s are largely unaffected by the growth of NFO layer. We would also like to highlight that the values of $n_i$'s and $\mu_i$'s for GAO(52nm)/STO fall within the same range as GAO(8nm)/STO and NFO(52nm)/GAO(8nm)/STO (see **Fig. S11**(c) and (d)).

Now, let us verify the contribution of AHE towards $R_{xy}(H)$ for our heterostructures at low temperatures. According to the two-band model, the Hall coefficient, $R_H = R_{xy}/\mu_0 H$ should follow a Lorentzian-like shape with a single minimum at zero field.[59,64] However, in presence of finite AHE, $R_H$ shows a prominent upturn close to the zero-field regime and thus deviates significantly from the Lorentzian-like shape.[59,64] As shown in **Fig. S9**(a), while $R_H$ at 10 K does not exhibit any upturn near the zero field for GAO(8nm)/STO, it shows a weak upturn below ±2.5 T for NFO(52nm)/GAO(8nm)/STO, and this upturn-like feature gets a bit more pronounced at 6 K (see **Fig. S9**(b)). Note that this upturn-like feature in $R_H$ originating from AHE is strongly temperature dependent and usually disappears at higher temperatures.[59] We found that this upturn-like feature in $R_H$ for NFO(52nm)/GAO(8nm)/STO is much weaker at 15 K compared to 10 K and entirely disappears above 15 K (see **Fig. S9**(c)). Therefore, while AHE contributes to $R_{xy}(H)$ in NFO(52nm)/GAO(8nm)/STO below 10 K, its contribution is suppressed significantly for $T \geq 15$ K, and hence the two-band conduction solely contributes to the observed non-linear Hall effect in this heterostructure in the temperature range: $15\text{ K} \leq T \leq 100$ K. In an attempt to determine the contribution of AHE, we fitted the $R_{xy}(H)$ data for NFO(52nm)/GAO(8nm)/STO at 6 K and 10 K using the combined two-band + AHE model (see **Fig. S10**), which is defined as,[59][64]

$$R_{xy}(\mu_0 H) = R_{xy}^{2e}(\mu_0 H) + R_{xy}^{AHE}(\mu_0 H) = R_{xy}^{2e}(\mu_0 H) + R_0^{AHE}\tanh(H/H_C) \qquad (4)$$

Here, the second term ($R_{xy}^{AHE}(\mu_0 H)$) represents the AHE contribution, wherein $R_0^{AHE}$ is the saturation value of $R_{xy}^{AHE}(\mu_0 H)$ when the external magnetic field attains the critical value $\mu_0 H = \mu_0 H_C$.[59][64] The comparison of the fitting parameters for the $R_{xy}(\mu_0 H)$ data fitted with only the two-band model and the combined two-band + AHE model for NFO(52nm)/GAO(8nm)/STO are shown in **Table S3** (**Table S4**) for $T$ = 6 K ($T$ = 10 K). It is evident that the contribution of AHE towards $R_{xy}(\mu_0 H)$ of NFO(52nm)/GAO(8nm)/STO is very small at 10 K but becomes significant at 6 K, which is consistent with the literature[59].

Let us now visualize the two-band electron transport from the point of view of the electronic band structure of the GAO(8nm)/STO interface. In the case of STO-based heterointerfaces, the $t_{2g}$ bands associated with the Ti $3d$ orbitals are responsible for interfacial electronic conduction.[66] Each $t_{2g}$ band is composed of three degenerate orbitals, namely, $d_{xy}$, $d_{xz}$ and $d_{yz}$. However, this degeneracy is lifted by inversion symmetry breaking when a 2DEG is formed at the interface between the $SrTiO_3$ surface and the top oxide layer (GAO, in our case).[67,68] Therefore, in STO-based 2DEGs, such interfacial inversion symmetry breaking leads to a perpendicular confining electric field at the interface that elevates the energy of the degenerate $d_{xz}/d_{yz}$ sub-bands with respect to the $d_{xy}$ sub band.[67,68] Typically, for the LAO/STO based 2DEGs with low carrier densities ($n < n_C = 1.7 \times 10^{13}$ cm$^{-2}$), only the $d_{xy}$ bands are occupied by the electrons.[66] However, when the carrier density exceeds the aforementioned critical limit ($i.e.$, $n \geq n_C = 1.7 \times 10^{13}$ cm$^{-2}$), the degenerate $d_{xz}/d_{yz}$ sub-bands start to be occupied.[66] Note that the aforementioned critical density of $n_C = 1.7 \times 10^{13}$ cm$^{-2}$ for the Lifshitz transition from one band to two band conduction was reported in case of LAO/STO based heterointerfaces.[66] On the other hand, different values of $n_C$ in the range between $2.6 \times 10^{13} - 3 \times 10^{13}$ cm$^{-2}$ have been reported in the case of GAO/STO based 2DEGs.[38,69] However, in our case, the total carrier density, $n_S = (n_1 + n_2) \approx 1.61 \times 10^{13}$ and $\approx 2.01 \times 10^{13}$ cm$^{-2}$ at $T = 100$ K and $n_S \approx 1.25 \times 10^{13}$ and $\approx 1.65 \times 10^{13}$ cm$^{-2}$ at $T = 10$ K for GAO(8nm)/STO and NFO(52nm)/GAO(8nm)/STO, respectively, and we have observed two band conduction for both the heterostructures below 100 K. Therefore, the critical carrier density, $n_C$ for GAO(8nm)/STO based 2DEGs can be considered to fall within the range: $1 \times 10^{13} \leq n_C \leq 3 \times 10^{13}$ cm$^{-2}$ and hence, both $d_{xy}$ and $d_{xz}/d_{yz}$ sub-bands associated with the $t_{2g}$ bands of the Ti $3d$ orbitals of STO participate in electronic transport in the GAO(8nm)/STO based 2DEGs,[10,38,70]. Nevertheless, the carrier density and electron mobility of the GAO/$SrTiO_3$ based 2DEG remains nearly unaffected by the growth of the NFO top layer.

## 2.4. Two-Terminal Electronic Transport

### *2.4.1. Temperature Dependence*

To acquire further insight into the charge transport behavior, we performed two-terminal current-voltage ($IV$) measurements on both heterostructures for $5\ \mathrm{K} \leq T \leq 60\ \mathrm{K}$. The heterostructures were mounted on a conventional PPMS resistivity puck. An Ag electrode of 1 mm diameter was used as the top electrode on the NFO (GAO) layer in NFO(52nm)/GAO(8nm)/STO (GAO(8nm)/STO) and an Al wire was ultrasonically wedge-bonded to the NFO (GAO) surface to make Ohmic contact to the 2DEG at the GAO(8nm)/STO interface, that served as the bottom electrode. A thin copper wire of 50 μm diameter was used to electrically connect the top Ag electrode to one of the Au bond pads of the PPMS puck and the Al wire connected to the 2DEG was wire-bonded to another Au pad of the puck. A similar two-terminal measurement geometry was also employed to study the $I-V$ characteristics of LAO/STO-based 2DEGs.[71,72] The $IV$ curves for GAO(8nm)/STO (NFO(52nm)/GAO(8nm)/STO) are linear and symmetric above 50 K (25 K) but become nonlinear as well as asymmetric for $T \leq 20$ K ($T \leq 50$ K) as shown in **Fig. 5**(a) (**Fig. 5**(b)) and the asymmetry gets stronger with further decreasing temperature. Nevertheless, the asymmetry of the $I-V$ curves is much weaker in GAO(8nm)/STO than in NFO(52nm)/GAO(8nm)/STO (see **Fig. 5**(f)). For example, in NFO(52nm)/GAO(8nm)/STO, the absolute value of the current, $|I|$ at $T = 5$ K increases rapidly with increasing positive bias voltage above ≈ +0.1 V and reaches ≈ 3.5 mA at +10 V, whereas, for the negative bias, $|I|$ increases very slowly from ≈ 10 nA at −0.1 V to ≈ 10 μA at −1 V and reaches ≈ 0.2 mA at −10 V. On the other hand, in GAO(8nm)/STO, $|I|$ at $T = 5$ K also increases rapidly with increasing positive bias voltage and reaches ≈ 7 mA at +10 V, whereas, for the negative bias, $|I|$ increases relatively slowly from ≈ 15 μA at −0.1 V to ≈ 0.25 mA at −1 V and reaches ≈ 3 mA at −1 V. Such asymmetric nonlinear $I-V$ curves resemble diode-like nonlinear rectifying behavior observed previously in STO-based 2DEGs,[71–77] and attributed to different mechanisms, *e.g.*, external electric field-induced tunneling of electrons from valence band of LAO to the STO potential-well,[72] accumulation of electron carriers in the 2DEG region under positive bias and depletion of the 2DEG under large negative bias[71,78], and Fermi-level pinning at the $Al_2O_3$/2DEG interface and the high Schottky barrier at the Pt/$Al_2O_3$ interface in Pt/$Al_2O_3$/2DEG/STO heterointerfaces stemming from asymmetric band alignment[76]. Notably, the $IV$ curves for both GAO(8nm)/STO and NFO(52nm)/GAO(8nm)/STO exhibit hysteresis in the reverse bias direction (see insets of **Figs. 5**(a) and (b), respectively), which is more pronounced and broader in NFO(52nm)/GAO(8nm)/STO compared to GAO(8nm)/STO and can be explained by the combined effect of electric field induced trapping/de-trapping of positively charged oxygen vacancies at the γ-$Al_2O_3$/$SrTiO_3$ interface[75] and migration of oxygen vacancies through the NFO layer[79,80] in NFO(52nm)/GAO(8nm)/STO.

**Figures 5**(c) and (d) display the temperature dependence of the absolute values of forward bias ($I_F$) and reverse bias ($I_R$) currents at the applied voltage of 10 V for GAO(8nm)/STO and NFO(52nm)/GAO(8nm)/STO, respectively. Clearly, $|I_F|$ and $|I_R|$ bifurcates from each other below 50 K (20 K) for NFO(52nm)/GAO(8nm)/STO (GAO(8nm)/STO). Additionally, $|I_F|$ and $|I_R|$ for both GAO(8nm)/STO and NFO(52nm)/GAO(8nm)/STO exhibit maxima in the vicinity of their corresponding Kondo-upturns, suggesting the contribution of the 2DEG towards the observed $I-V$ characteristics[71]. The rectification ratio, defined as, $\left|\frac{I_F(+10\text{ V})}{I_R(-10\text{ V})}\right|$ increases dramatically with decreasing temperature for both the heterostructures especially below their respective Kondo-upturns (see

Fig. 5(e)), and is nearly an order of magnitude higher in NFO(52nm)/GAO(8nm)/STO than in GAO(8nm)/STO at 5 K, which underscores the higher performance of NFO(52nm)/GAO(8nm)/STO as a rectifier. We verified these observations for another two-terminal contact configuration for both GAO(8nm)/STO and NFO(52nm)/GAO(8nm)/STO (see **Figure S14**) along with the robustness of the rectification performance of NFO(52nm)/GAO(8nm)/STO at 5 K (see **Figure S15**).

In order to verify if the thickness of insulating oxide layers plays a dominant role towards the rectification performance of two-terminal devices, we have further performed two-terminal $I - V$ measurements at different temperatures between 5-100 K on GAO(52nm)/STO using the same measurement conditions as for NFO(52nm)/GAO(8nm)/STO and GAO(8nm)/STO (see **Fig. S17**). We found that the $IV$ curves for Ag/GAO(52nm)/STO are also asymmetric and nonlinear and show diode-like rectifying behavior below $T \leq 75$ K and the asymmetry gets stronger with further decreasing temperature (see **Fig. S17(a)**). Notably, the asymmetry of the $I - V$ curves is stronger in Ag/GAO(52nm)/STO than in Ag/GAO(8nm)/STO. Furthermore, $|I_F|$ and $|I_R|$ bifurcates from each other below 100 K for GAO(52nm)/STO, and both of them exhibit maxima in the vicinity of the corresponding Kondo-upturn (see **Fig. S17(b)**), similar to what was observed in NFO(52nm)/GAO(8nm)/STO and GAO(8nm)/STO. The rectification ratio, $\left|\frac{I_F(+10\text{ V})}{I_R(-10\text{ V})}\right|$ increases with decreasing temperature for GAO(52nm)/STO and shows a more rapid enhancement below the Kondo-upturn (see **Fig. S17(c)**). This is important to note that $\left|\frac{I_F(+10\text{ V})}{I_R(-10\text{ V})}\right|$ for GAO(52nm)/STO is nearly the same as in GAO(8nm)/STO at $T = 5$ K but an order of magnitude lower than that of NFO(52nm)/GAO(8nm)/STO which highlights the higher performance of NFO(52nm)/GAO(8nm)/STO as a diode-like rectifier than both GAO(8nm)/STO and GAO(52nm)/STO films. This observation also indicates that the total thickness of the insulating oxide layers does not govern the rectification performance of these heterostructures.

*2.4.2. Magnetic Field Dependence*

**Figures 6**(a) and (b) show two-terminal *IV* curves measured at 5 K under different out-of-plane magnetic fields in the range $0\,\mathrm{T} \leq \mu_0 H \leq 9\,\mathrm{T}$ for GAO(8nm)/STO and NFO(52nm)/GAO(8nm)/STO, respectively. Clearly, both $|I_F|$ and $|I_R|$ for GAO(8nm)/STO and NFO(52nm)/GAO(8nm)/STO decrease with increasing magnetic field strength at the applied voltage of 10 V, which is also evident from **Figs. 6**(e) and (f). However, the relative changes in the magnetic field dependences of both $|I_F|$ and $|I_R|$ – defined as $\left(\frac{\Delta I_F}{I_F} = [\{I_F(\mu_0 H) - I_F(\mu_0 H = 0)\}/I_F(\mu_0 H = 0)] \times 100\%\right)$ and $\left(\frac{\Delta I_R}{I_R} = [\{I_R(\mu_0 H) - I_R(\mu_0 H = 0)\}/I_R(\mu_0 H = 0)] \times 100\%\right)$, respectively at 10 K and in the field range: $-9\mathrm{T} \leq \mu_0 H \leq +9\mathrm{T}$ are lower in GAO(8nm)/STO compared to NFO(52nm)/GAO(8nm)/STO (see **Figs. 6**(e) and (f), respectively). On the contrary, the magnetic field dependence of $|I_F|$ and $|I_R|$ at 10 V for GAO(52nm)/STO shows distinct behavior: $|I_F|$ increases with magnetic field strength and $|I_R|$ shows a decrease with increasing the magnetic field strength for GAO(52nm)/STO at the applied voltage of 10 V, as shown in **Figs. S18**(a) and (b). This contrasting behavior of $|I_F|$ and $|I_R|$ at 10 V could be due to the anomalous feature observed in the $I - V$ curves at 5 K in the applied voltage range: $+8.5\,\mathrm{V} \leq V \leq +10\,\mathrm{V}$ (see **Figs. S18**(a)) Interestingly, both $|I_F|$ and $|I_R|$ at 5 K for GAO(52nm)/STO decrease with increasing magnetic field strength for the applied voltage of 8 V (see **Figs. S18**(c)), which is consistent with those observed in NFO(52nm)/GAO(8nm)/STO and GAO(8nm)/STO.

Most importantly, the rectification ratio, $\left|\frac{I_F(+10\text{ V})}{I_R(-10\text{ V})}\right|$ for NFO(52nm)/GAO(8nm)/STO increases significantly with increasing magnetic field (see **Fig. 6**(d)) whereas that for GAO(8nm)/STO shows a small decrease with increasing magnetic field (see **Fig. 6**(c)). We observed similar magnetic field dependences of the $I - V$ curves across another two-terminal contact configuration for both GAO(8nm)/STO and NFO(52nm)/GAO(8nm)/STO (see **Figure S16**). In case of GAO(8nm)/STO, the rectification ratio at 10 V decreases from 2.3 at 0 T to 2.2 at 9 T with a magneto-electronic rectification effect, defined as, $MER(\%) = \left[\left\{\left|\frac{I_F(+10\text{ V})}{I_R(-10\text{ V})}\right|_{\mu_0 H=9\text{ T}} - \left|\frac{I_F(+10\text{ V})}{I_R(-10\text{ V})}\right|_{\mu_0 H=0\text{ T}}\right\} / \left|\frac{I_F(+10\text{ V})}{I_R(-10\text{ V})}\right|_{\mu_0 H=0\text{ T}}\right] \times 100\%$ of -2.5% at 5 K. On the other hand, for NFO(52nm)/GAO(8nm)/STO, the rectification ratio at 10 V increases from 18.6 at 0 T to 34.6 at 9 T with $MER = +86.2\%$. However, for both 8 V and 10 V applied bias voltages, the rectification ratio for GAO(52nm)/STO showed a positive magnetic field dependence similar to NFO(52nm)/GAO(8nm)/STO but unlike GAO(8nm)/STO. For GAO(52nm)/STO, the rectification ratio at 10 V (8 V) increases from 3.22 (4.9) at 0 T to 4.4 (5.3) at 9 T with $MER = +36.6\%$ (8.5%) (see **Figs. S18**(d) and (e), respectively). Therefore, NFO(52nm)/GAO(8nm)/STO shows the highest $MER$ at 5 K among all the three samples, which underscores higher performance of NFO(52nm)/GAO(8nm)/STO as a magnetic diode,[81–84] with large positive magneto-electronic rectification effect.

*2.4.3. Possible Origin of the Rectification Effect*

As observed in the preceding sections, NFO(52nm)/GAO(8nm)/STO shows much higher rectification performance than both GAO(8nm)/STO and GAO(52nm)/STO at $T = 5\text{ K}$ albeit the fact that the total thickness of the insulating oxide layers in NFO(52nm)/GAO(8nm)/STO is close to that in GAO(52nm)/STO. Therefore, the electric-field-induced electron tunneling between the Ag electrode and the 2DEG can be ruled out as a possible mechanism for the observed nonlinear rectification effect. Instead, electric field induced migration of ions like positively charged oxygen vacancies inside the GAO and (GAO+NFO) layer(s) can be considered as a possible origin of the observed nonlinear rectification effect in the GAO(8nm)/STO and NFO(52nm)/GAO(8nm)/STO heterostructures, as also observed in Pt/$Al_2O_3$/$SrTiO_3$ based 2DEG systems.[76] Such electric field induced motion of the positively charged oxygen vacancies can result in locally varying electronic properties in oxide thin films (in our case, NFO and GAO),[85–87] which could be the origin of such enhanced through-plane electronic currents in GAO(8nm)/STO and NFO(52nm)/GAO(8nm)/STO. We would also like to highlight that our magnetometry results for NFO(52nm)/GAO(8nm)/STO support the presence of oxygen vacancies in the NFO layer. However, for a more accurate mechanistic understanding of the observed diode-like rectifying behavior in these heterostructures, a systematic study is needed. Specifically, a systematic variation of the top electrode material (*e.g.*, Al, Au, Cu, Pt, etc.) and size, in combination with a detailed analysis based on the existing fundamental concepts on valence change mechanisms in oxide electronic devices,[88] would be worth investigating in the future.

**3. Conclusion**

To summarize, we have demonstrated that a ferrimagnetic insulating singly oriented crystalline NFO(001) film can be grown by RF sputtering on top of GAO(001)/STO(001) at 150 ℃ without compromising the mobility and charge carrier density of the 2DEG at the GAO(8nm)/STO interface. While the GAO(8nm)/STO interface was found to be nonmagnetic, NFO(52nm)/GAO(8nm)/STO showed well-defined magnetic hysteresis loops at room temperature as well as at low temperatures. Moreover, NFO(52nm)/GAO(8nm)/STO showed a glassy magnetic ground state and considerable exchange bias effect at low temperatures, confirming the presence of oxygen vacancies in the NFO film. The sheet resistance of both heterostructures exhibited metallic behavior throughout the measured temperature range, with low temperature resistivity upturns driven by the combined effect of Kondo-like scattering and weak antilocalization. Furthermore, we have shown that NFO(52nm)/GAO(8nm)/STO behaves as a magnetic diode with positive magneto-electronic rectification effect at low temperatures, which will be beneficial for the development of nanoscale all-oxide heterostructures capable of efficient spin-charge conversion. Electric field induced migration of positively charged oxygen vacancies can be considered a possible origin of the observed nonlinear diode-like rectifying behavior in these heterostructures.

## 4. Experimental Section/Methods

### 4.1. Thin Film Growth

High quality GAO(001) thin films were grown on $TiO_2$-terminated single crystalline STO(001) substrates by pulsed laser deposition (PLD) using a KrF excimer laser ($\lambda = 248$ nm) and a single crystalline $\alpha$-$Al_2O_3$ target at 550° C with an $O_2$ partial pressure of $1 \times 10^{-6}$ mbar, laser fluence of 8 J/cm$^2$, a repetition rate of 1 Hz, and the substrate-to-target distance of 4.5 cm.[10,11] NFO thin film was subsequently deposited *ex-situ* on the PLD-grown GAO(001)/STO(001) by a reactive RF magnetron sputtering with a commercial polycrystalline NFO target. During sputtering, the substrate-target distance was kept to 6 cm, and the substrate temperature was fixed at 150°C. The base pressure of the sputtering chamber was maintained at $1 \times 10^{-6}$ mbar. A mixture of Ar and $O_2$ gases with 5:1 ratio was used during sputtering with a total working pressure of ~$3 \times 10^{-2}$ mbar and an RF power of 150 W.

### 4.2. Structural, morphological and magnetic characterization

Structural properties of the GAO/STO and NFO(52nm)/GAO(8nm)/STO films were characterized with X-ray diffraction (XRD) and reflectivity (XRR) measurements using the Rigaku SmartLab diffractometer and morphological properties of these heterostructures were characterized with Bruker Atomic Force Microscopy (AFM). Magnetic properties of these heterostructures were characterized using vibrating sample magnetometry (VSM) associated with the physical property measurement system (PPMS), Quantum Design.

**4.3. X-ray photoelectron spectroscopy**

X-ray photoelectron spectroscopy (XPS) measurements were performed using a Thermo Fisher Nexsa system equipped with an Al K-alpha X-ray source. To minimize adventitious carbon contamination, the sample surface was subjected to low-energy argon cluster etching (2000 eV, cluster size 2000) for 500 s. All spectra were processed in CasaXPS and referenced to the C 1s peak at 285 eV. The high-resolution spectra were smoothed using a moving average algorithm with a width of 5 data points.

**4.4. Electrical Transport Measurements**

Four-terminal electronic transport measurements (sheet resistance and Hall measurements) on these heterostructures were performed using a cryogen-free low temperature measurement system equipped with a 16 T magnet as well as the DC resistivity option of the PPMS. In the cryostat, the samples were mounted on ceramic leadless 20-pin chip carriers and electrical connections on these samples were made by Al ultrasonic wedge-type wire-bonding in four-point van-der-Pauw (vdP) configuration. A Keithley 2400 was used for sourcing DC current and a Keithley 2182a nanovoltmeter was used to read the voltage. Temperature and magnetic field dependence of the two-terminal current-voltage measurements were performed using the PPMS. The average distance between the Ag electrode and the Al wire along the film surface was approximately 1 mm for both NFO(52nm)/GAO(8nm)/STO and GAO(8nm)/STO. A DC bias voltage was applied to the top Ag electrode and swept from $+10\ \mathrm{V} \rightarrow -10\ \mathrm{V} \rightarrow +10\ \mathrm{V}$ with a sweep-rate of 0.1 V/s and, the current flowing through the entire circuit was measured by a Keithley 6487 pico-ammeter.

**Acknowledgements**

A. C., T. H. O., A. G., and F. T. acknowledge support by research grants 37338 (SANSIT) and 69171 (ETHOS) from Villum Fonden. We are thankful to Roxy Lee for her assistance in the interpretation of the XPS spectra and the quantification of the XPS data.

**Data Availability Statement**

The data that support the findings of this study are available from the corresponding author upon reasonable request.

**List of Figures**

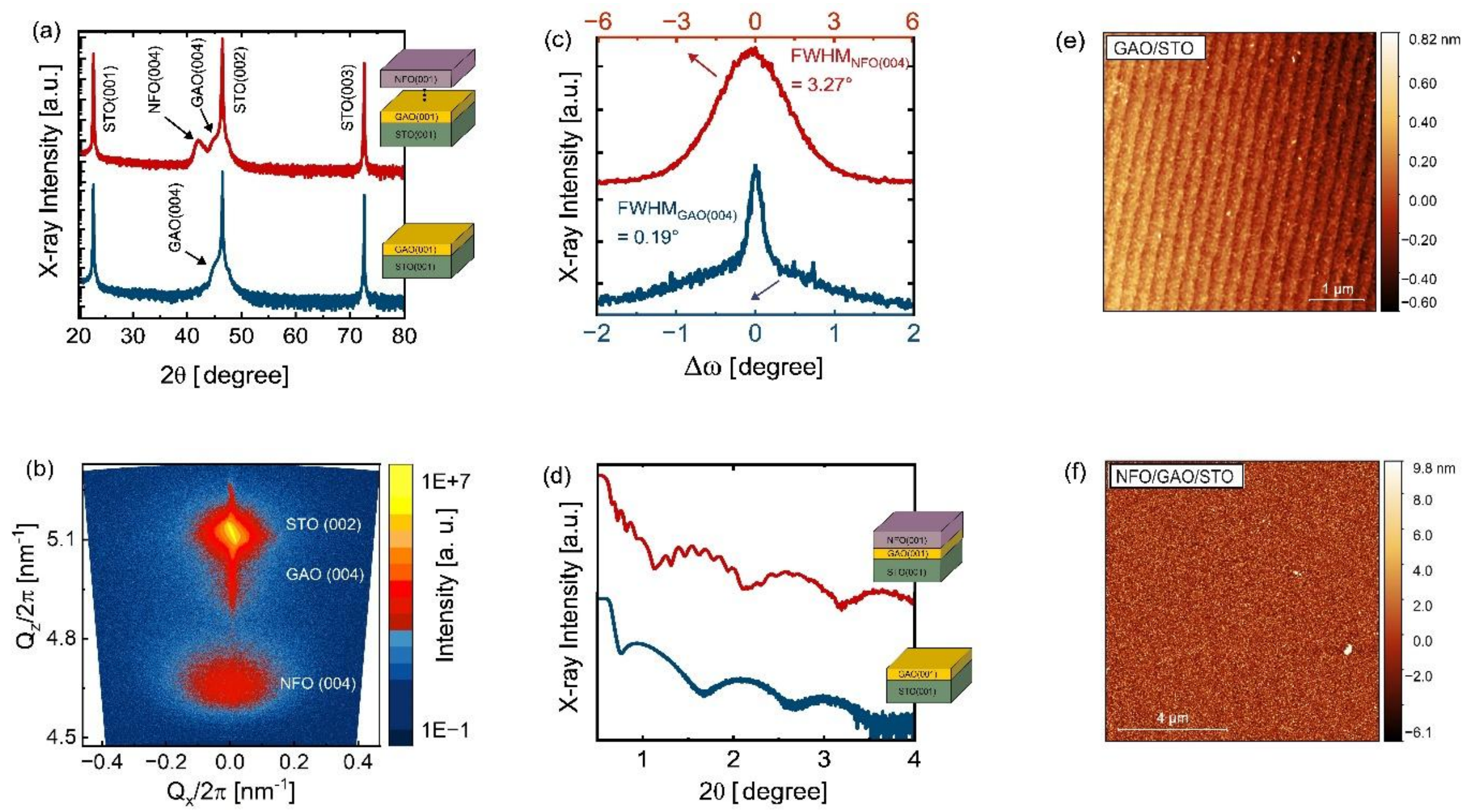

**Figure 1.** (a) X-ray diffraction (XRD) patterns for the GAO(8nm)/STO and NFO(52nm)/GAO(8nm)/STO films recorded at room temperature using a Rigaku SmartLab diffractometer, (b) reciprocal-space-mapping performed on the NFO(52nm)/GAO(8nm)/STO heterostructure around the symmetric (002) peak of STO shows distinct peaks associated with the (004) planes of both GAO and NFO, (c) the X-ray rocking curves recorded on NFO(52nm)/GAO(8nm)/STO for the (004)-reflections of both NFO and GAO suggesting higher crystallinity of GAO with a full-width-at-half-maximum (FWHM) of $(0.2 \pm 0.01)^{\circ}$ compared to NFO with an FWHM of $(3.27 \pm 0.01)^{\circ}$, and (d) the Kiessig fringes in the X-ray reflectivity (XRR) data for GAO(8nm)/STO and NFO(52nm)/GAO(8nm)/STO. Atomic force microscopy (AFM) images of (e) GAO(8nm)/STO with root-mean-squares (RMS) roughness of 0.5 nm and (f) NFO(52nm)/GAO(8nm)/STO with RMS roughness of 2 nm.

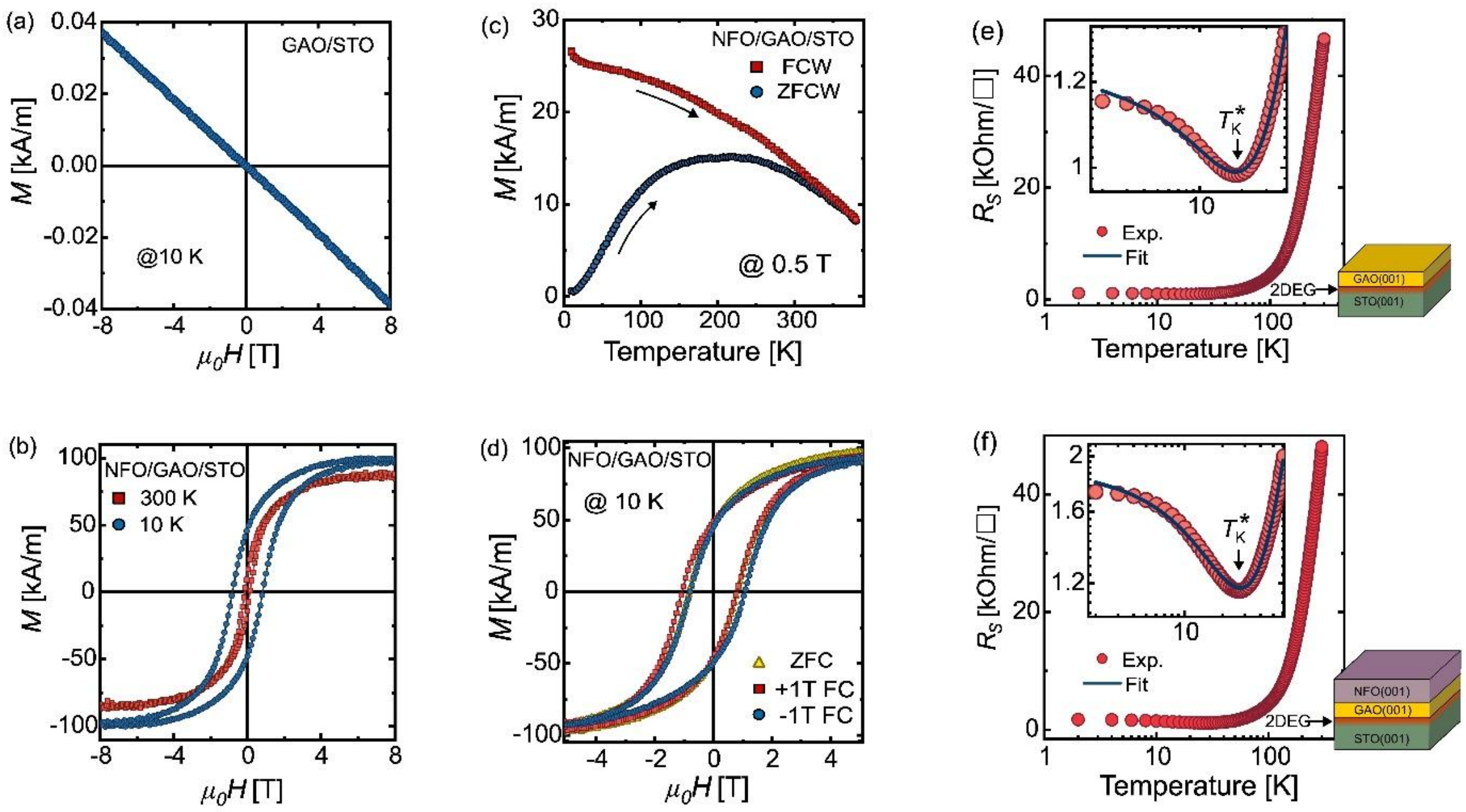

**Figure 2.** (a) Magnetic field dependence of magnetization, $M(H)$ for GAO(8nm)/STO measured while sweeping an in-plane magnetic field between $\mu_0 H = \pm 9$ T at 10 K showing diamagnetic behavior of GAO(8nm)/STO, (b) $M(H)$ hysteresis loops for NFO(52nm)/GAO(8nm)/STO at 300 and 10 K, (c) temperature dependence of magnetization, $M(T)$ of NFO(52nm)/GAO(8nm)/STO measured under the zero-field cooled warming (ZFCW) and field cooled warming (FCW) protocols in the temperature ($T$) range: $10\text{ K} \leq T \leq 390\text{ K}$ in presence of an in-plane magnetic field of $\mu_0 H = 0.5$ T, (d) Exchange bias effect: $M(H)$ loops for NFO(52nm)/GAO(8nm)/STO measured at 10 K with zero field cooled and field cooled conditions with cooling fields of $\mu_0 H = +1$ T and $-1$ T. (e) and (f) Main panels: temperature dependence of the sheet resistance, $R_S(T)$ for the GAO(8nm)/STO and NFO(52nm)/GAO(8nm)/STO heterostructures, respectively, measured in temperature the range: $2\text{ K} \leq T \leq 300\text{ K}$, insets show the fitting of the low temperature upturns in $R_S(T)$ using the Kondo model.

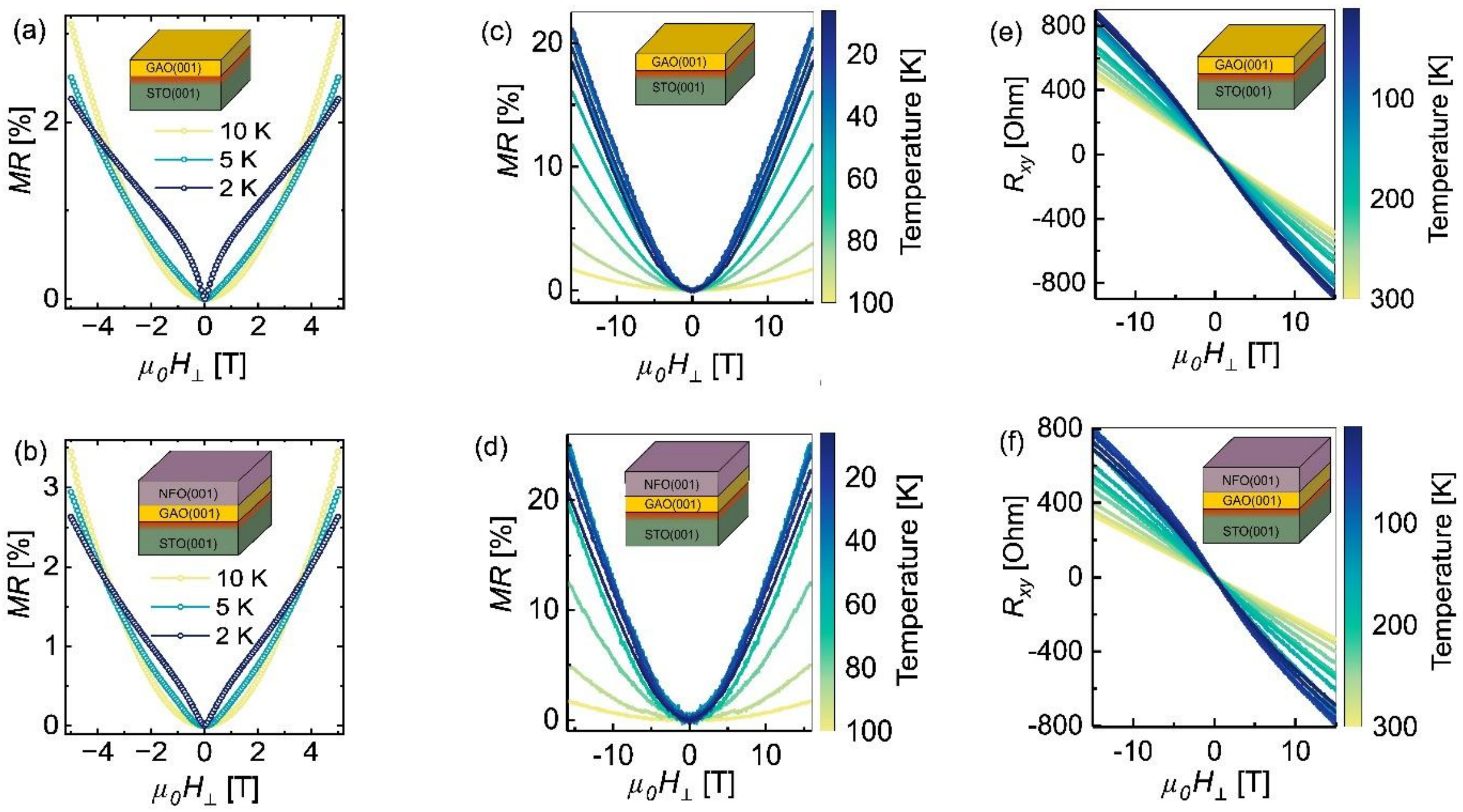

**Figure 3.** Magnetic field dependence of magnetoresistance (MR), defined as, $MR = [\{R_s(\mu_0 H) - R_s(\mu_0 H = 0)\}/R_s(\mu_0 H = 0)]$ for (a) GAO(8nm)/STO and (b) NFO(52nm)/GAO(8nm)/STO while sweeping an out-of-plane magnetic field between $\mu_0 H = \pm 5$ T at 2 K, 5 K and 10 K. Magnetic field dependence of MR for (c) GAO(8nm)/STO and (d) NFO(52nm)/GAO(8nm)/STO while sweeping an out-of-plane magnetic field up to $\mu_0 H = \pm 16$ T in the temperature range $6\,K \leq T \leq 100\,K$. Magnetic field dependence of Hall resistance, $R_{xy}(H)$ at different temperatures in the range $10\text{ K} \leq T \leq 300\text{ K}$ with externally applied magnetic fields up to $\mu_0 H = \pm 15$ T for (e) GAO(8nm)/STO and (f) NFO(52nm)/GAO(8nm)/STO.

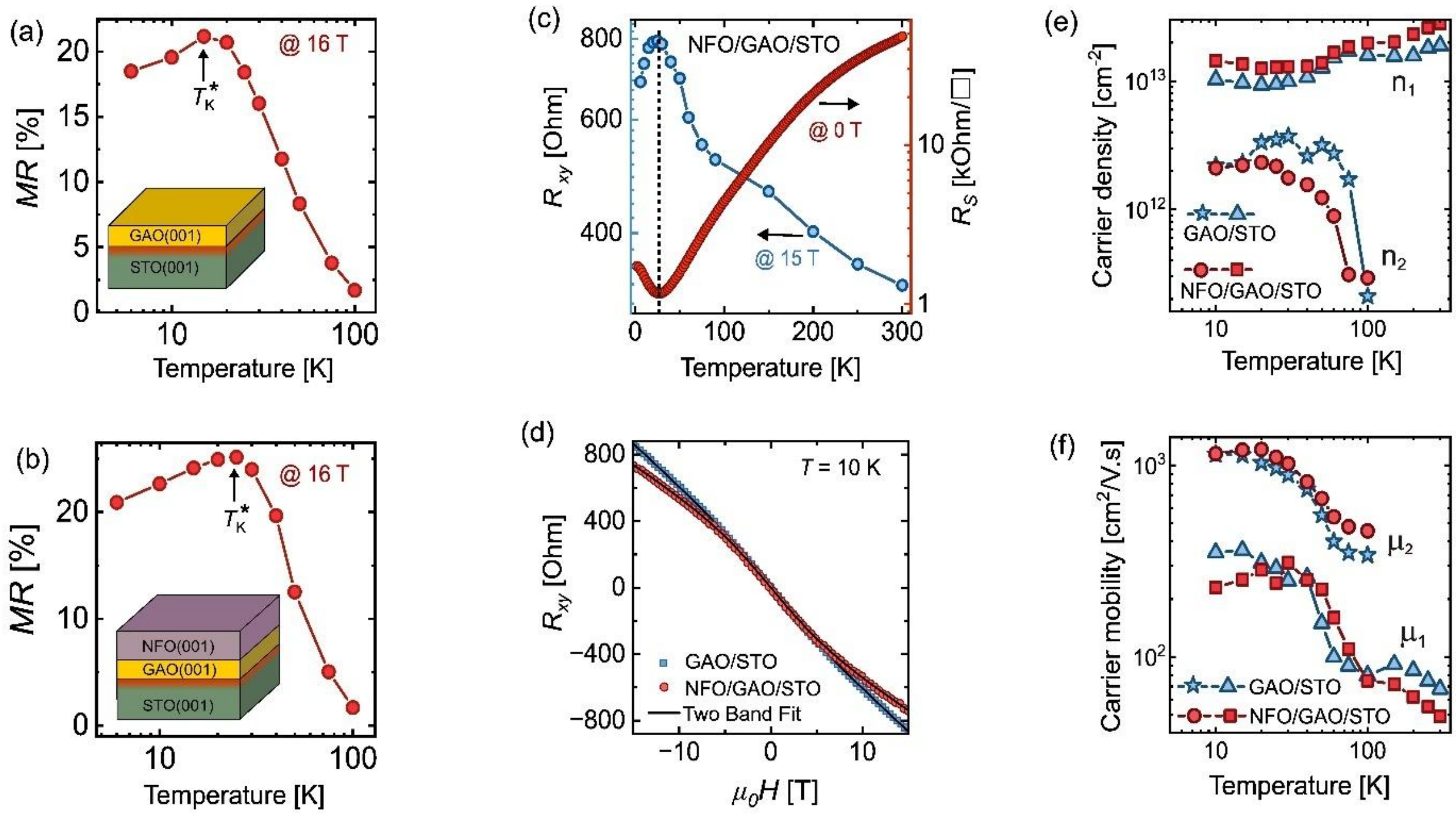

**Figure 4.** Temperature dependence of magnetoresistance, $MR = [\{R_s(\mu_0 H) - R_s(\mu_0 H = 0)\}/R_s(\mu_0 H = 0)]$ at $\mu_0 H = 16$ T for (a) GAO(8nm)/STO and (b) NFO(52nm)/GAO(8nm)/STO. (c) Left vertical axis: Temperature dependence of $R_{xy}$ at $\mu_0 H = +15$ T for NFO(52nm)/GAO(8nm)/STO and right vertical axis: Corresponding $R_S(T)$ for NFO(52nm)/GAO(8nm)/STO. (d) The $R_{xy}(H)$ curves for both GAO(8nm)/STO and NFO(52nm)/GAO(8nm)/STO at 10 K fitted by the two-band conduction model using Eqns. (2) and (3). Temperature evolutions of (e) the carrier densities, $n_i$'s and (f) carrier mobilities, $\mu_i$'s for both GAO(8nm)/STO and NFO(52nm)/GAO(8nm)/STO.

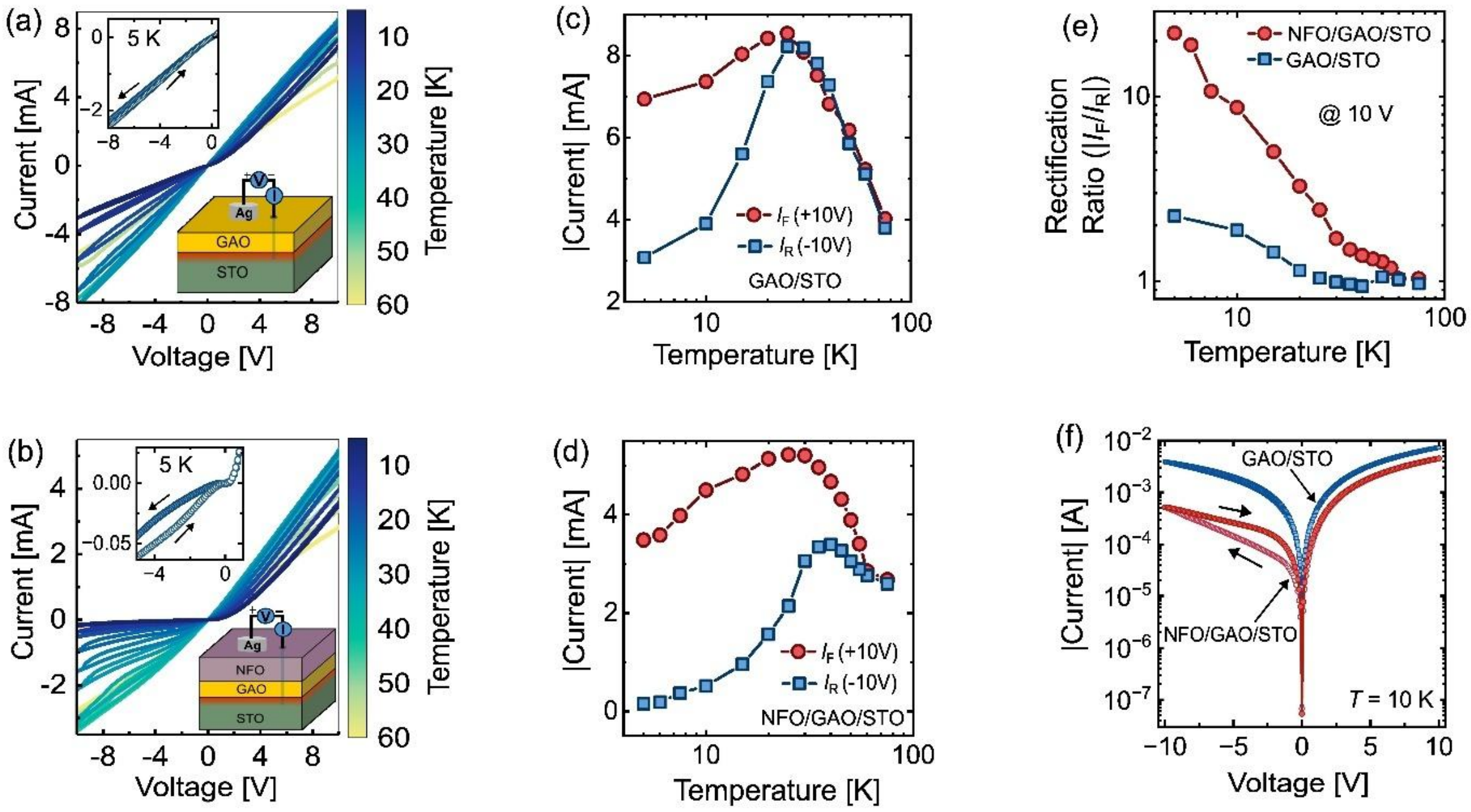

**Figure 5.** Main panels: Two-terminal $IV$ curves measured at different temperatures in the range: $5\text{ K} \leq T \ \leq 60\text{ K}$ while sweeping the DC bias voltage from $+10\text{ V} \rightarrow -10\text{ V} \rightarrow +10\text{ V}$ for (a) GAO(8nm)/STO and (b) NFO(52nm)/GAO(8nm)/STO, respectively, insets show zoomed-in view of the corresponding two-terminal $IV$ curves at 5 K in the reverse bias region. Temperature dependence of the absolute values of the forward bias ($I_F$) and reverse bias ($I_R$) currents at the applied voltage of 10 V for (c) GAO(8nm)/STO and (d) NFO(52nm)/GAO(8nm)/STO. (e) Temperature dependence of the rectification ratio for GAO(8nm)/STO and NFO(52nm)/GAO(8nm)/STO measured at the applied voltage of 10 V. (f) Comparison of the logarithmic $IV$ plots measured at 10 K for NFO(52nm)/GAO(8nm)/STO and GAO(8nm)/STO.

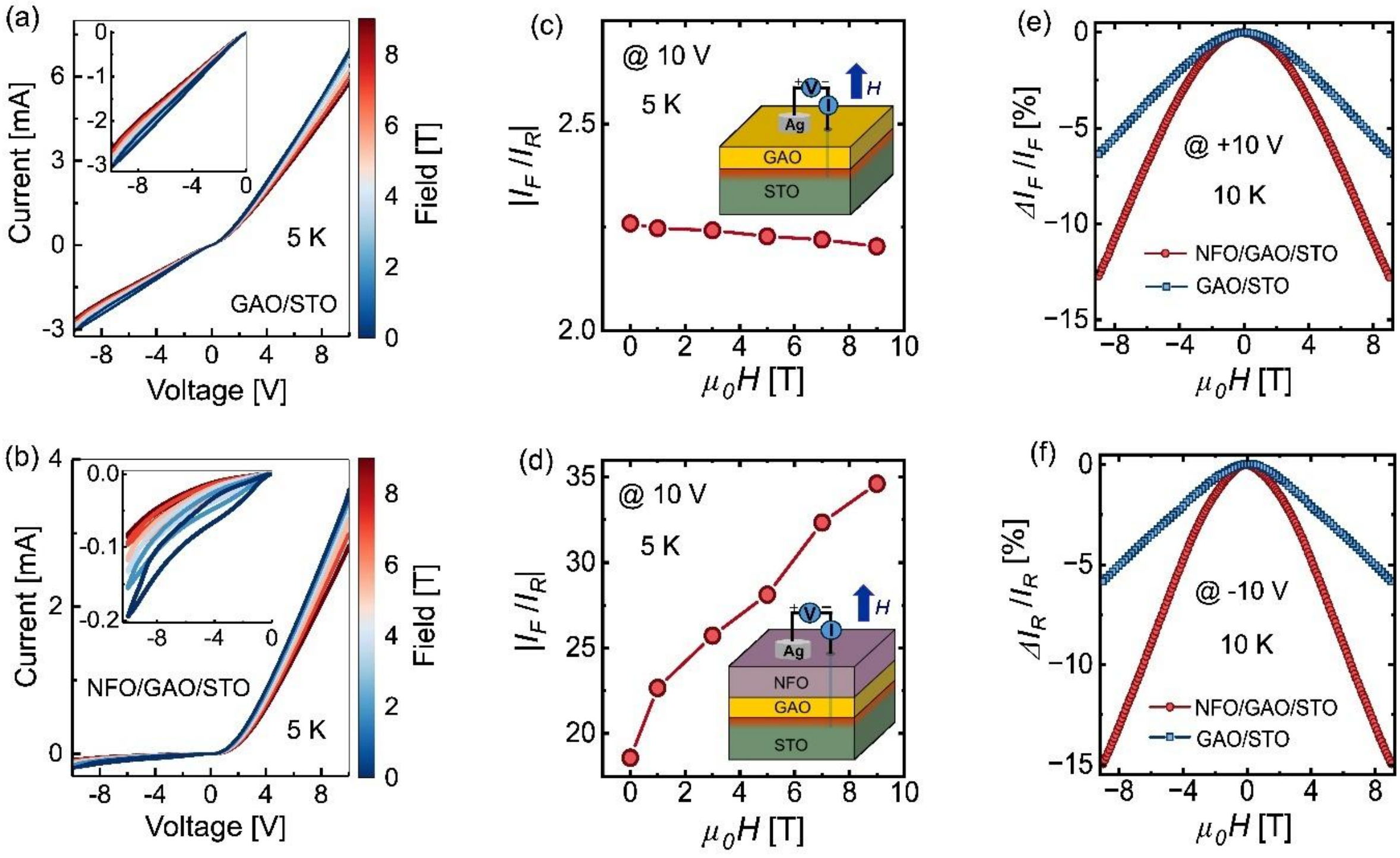

**Figure 6.** Main panels: two-terminal *IV* curves measured at $T = 5$ K in the presence of different magnetic fields in the range: $0\ \mathrm{T} \leq \mu_0 H \leq 9$ T while sweeping the DC voltage from $+10\ \mathrm{V} \rightarrow -10\ \mathrm{V} \rightarrow +10\ \mathrm{V}$ for (a) GAO(8nm)/STO and (b) NFO(52nm)/GAO(8nm)/STO, insets show zoomed-in view of the corresponding two-terminal *IV* curves in the reverse bias region. The corresponding magnetic field dependence of the rectification ratio for (c) GAO(8nm)/STO and (d) NFO(52nm)/GAO(8nm)/STO, respectively at the applied voltage of 10 V. Percentage changes in the magnetic field dependences of (e) $|I_F|$ and (f) $|I_R|$ for recorded at the applied voltages of +10 V and -10 V, respectively while sweeping the out-of-plane magnetic field at 10 K for both NFO(52nm)/GAO(8nm)/STO and GAO(8nm)/STO.